\documentclass[aps,showpacs,preprint] {revtex4}
\usepackage{graphicx}
\usepackage{bm}
\usepackage{amsfonts,amsmath,amssymb}
\usepackage[squaren]{SIunits}
\makeatletter
\begin{document}

\title{Effect of  iron-doping on  spin-state transition and  ferromagnetism in Pr$_{0.5}$Ca$_{0.5}$CoO$_{3-\delta}$ cobalt oxides}
\author{X. G. Luo, X. Li, G. Y. Wang, G. Wu}
\author{X. H. Chen}
\altaffiliation{Corresponding author}
\email{chenxh@ustc.edu.cn}\affiliation{Hefei National Laboratory for
Physical Science at Microscale and Department of Physics, University
of Science and Technology of China, Hefei, Anhui 230026, People's
Republic of China}

\begin{abstract}
Resistivity and dc magnetization measurements were performed for
the polycrystalline
Pr$_{0.5}$Ca$_{0.5}$Co$_{1-x}$Fe$_{x}$O$_{3-\delta}$ ($x$ = 0,
0.05, 0.10 and 0.15) samples. The as-fabricated samples exhibit
ferromagnetic (FM) transition and the transition temperature
increases with increasing the iron doping level. Annealing under
high oxygen pressure induces a spin-state transition of Co ions in
the \emph{iron-free} sample and such transition is reinforced with
increasing the annealing oxygen pressure, while the annealing
under high oxygen pressure suppresses the ferromagnetic ordering.
Contrary to the case of the \emph{iron-free} sample, no spin-state
transition is induced by the annealing under high oxygen pressure
for the \emph{iron-doped} samples, and the ferromagnetic
transition temperature is nearly independent of the annealing
procedures. The enhancement of the spin-state transition in the
\emph{iron-free} sample after annealing under high oxygen pressure
should be attributed to the reduction of the cell volume. The
suppression of the spin-state transition by the Fe doping is
related to the enlargement of the cell volume and the stronger
Fe-O bonds than Co-O bonds. The enhancement of the ferromagnetism
by the iron-doping might arise from the ferromagnetic exchange
interaction between Fe$^{3+}$ and Co$^{4+}$ through oxygen
(Fe$^{3+}$-O-Co$^{4+}$).

\end{abstract}


\maketitle

\section{INTRODUCTION}
An intriguing feature of the perovskite-related cobalt oxides is the
existence of spin-state transition (SST), which can be induced by
temperature, pressure, or carrier concentration. There are various
spin states for trivalent (low-spin LS: $t_{\rm 2g}^6e_{g}^0$;
intermediate-spin IS: $t_{2g}^5e_{g}^1$; high-spin HS:
$t_{2g}^4e_{g}^2$) and tetravalent cobalt ions (LS:
$t_{2g}^5e_{g}^0$; IS: $t_{2g}^4e_{g}^1$; HS: $t_{2g}^3e_{g}^2$). In
the simplest perovskite-type LnCoO$_3$ (Ln=Y, and rare earth
elements), a transition from low-spin to high- or intermediate-spin
was observed with increasing the
temperature.\cite{Yamaguchi,Sudheendra,Yan,Hejtmanek,Kzinek} In
another interesting oxygen-deficient cobalt oxides
LnBaCo$_2$O$_{5+\delta}$ (Ln= Pr, Nd, Sm, Eu, Gd and Tb), an abrupt
SST was reported as $\delta$ =
0.5.\cite{Kasper,Moritomo,Akahoshi,Frontera,FFrontera} The
conversion of different spin-state arises from the competition
between comparable the crystal-field with energy $\Delta_{\rm CF}$
($t_{\rm 2g}$-$e_{\rm g}$ splitting) and the intraatomic (Hund)
exchange with energy $J_{\rm ex}$ in magnitude, leading to
redistribution of electrons between $t_{\rm 2g}$ and $e_{\rm g}$
levels. $\Delta_{\rm CF}$ is found to be very sensitive to the
variation in the Co-O bond length ($d_{\rm Co-O}$), so that the
subtle balance between $\Delta_{\rm CF}$ and $J_{\rm ex}$ may be
easily disrupted by different kind of effects such as hole-doping
and chemical or external pressure.\cite{Asai,Vogt,Lengsdorf,Fita}
Very recently, Lengsdorf er al. and Fita et al. reported that the
pressure suppressed FM interactions in La$_{0.82}$Sr$_{0.18}$CoO$_3$
and La$_{1-x}$Ca$_{x}$CoO$_3$ due to the reduction of the population
of the $e_g$ electrons induced by the increase in the energy of
crystal field splitting.\cite{Lengsdorf,Fita}

The transition of the spin state of cobalt ions is always connected
with the change of conductivity. The metal-insulator transition
(MIT) is usually accompanied with a SST. A metal to insulator
transition in resistivity occurs in LaCoO$_3$ as the spin-state
changes gradually from high-spin to low-spin state between 350 K to
110 K.\cite{Goodenough} In LnBaCo$_2$O$_{5+\delta}$ (Ln= Pr, Nd, Sm,
Eu, Gd and Tb), an abrupt MIT takes place simultaneously as the SST
occurs.\cite{Kasper,Moritomo,Akahoshi,Frontera,FFrontera} The
occurrence of MIT accompanying with the SST suggests the essential
role of the $e_g$ electrons in the metallic electronic conduction.

In the present paper, we paid our attention on another cobalt
oxide Pr$_{0.5}$Ca$_{0.5}$CoO$_{3-\delta}$, which was reported
firstly by Tsubouchi et al.\cite{Tsubouchi1,Tsubouchi2} to exhibit
the simultaneous MIT and SST at low temperature. They revealed
that the MIT and SST in Pr$_{0.5}$Ca$_{0.5}$CoO$_{3-\delta}$ arise
from the large decrease of Co-O-Co angle around the transition
temperature with decreasing temperature. The large decrease of
Co-O-Co angle results in an extraction of lattice volume and a
reduction in the covalence of the Co-O bonds, which enlarges the
splitting of the crystal field and stabilizes the localized
low-spin state. As one knows, the ferromagnetic exchanges occur
mainly through the $e_{\rm g}$ hopping in perovskite-type cobalt
oxides. The reduction in the population of the $e_{\rm g}$
electrons has an effect of suppressing ferromagnetism, which has
been demonstrated by previous researches of using external
pressure \cite{Lengsdorf,Fita}or chemical
pressure.\cite{Kriener,Luo} Therefore, the degradation of
spin-state of cobalt ions would suppress the $e_{\rm g}$ hopping
and consequently the ferromagnetic interaction. Actually,
Tsubouchi et al. \cite{Tsubouchi1,Tsubouchi2} presented that the
long ferromagnetic order does not appear down to 10 K in
Pr$_{0.5}$Ca$_{0.5}$CoO$_3$. It seems to suggest that the
spin-state transition is competitive with the ferromagnetism.
Thus, if ferromagnetism is enhanced through some ion substitution,
spin-state transition is expected to be suppressed. Maignan et al.
proposed that the superexchange between the high-spin Fe$^{3+}$
and the low-spin Co$^{4+}$ ions through the oxygen atom could be
ferromagnetic.\cite{Maignan} In this work, we examined effect of
Fe doping and annealing procedure under the high oxygen pressure
(HP) on the ferromagnetism, spin-state transition, and charge
transport in the polycrystalline
Pr$_{0.5}$Ca$_{0.5}$Co$_{1-x}$Fe$_{x}$O$_{3-\delta}$ samples. The
results indicate that the ferromagnetism is enhanced, while
spin-state transition is suppressed by the Fe doping. On the other
hand, the annealing procedures under the high oxygen pressure
induce a spin-state transition in the \emph{iron-free} sample, but
do not change the ferromagnetic transition temperature in the
iron-doped samples.

\section{EXPERIMENTAL PROCEDURES}
The polycrystalline ceramic
Pr$_{0.5}$Ca$_{0.5}$Co$_{1-x}$Fe$_x$O$_{3-\delta}$ samples with
$x$ = 0, 0.05, 0.10 and 0.15 were synthesized using conventional
solid state reaction meathod. They were prepared from a
stoichiometric mixture of the fine powder of oxides:
Pr$_6$O$_{11}$, CaCO$_3$, Co$_2$O$_3$ and Fe$_2$O$_3$. The
stoichiometric mixture was ground carefully until the mixture
became homogeneous. The mixtures then were calcined at
1200$\celsius$ in the flowing oxygen for 24 h. The calcined powder
was reground and pressed into pellets. The pellets were sintered
at 1200$\celsius$ in the flowing oxygen for 24 h and finally
cooled to room temperature. Some portions of pellets for each
composition were annealed at 600$\celsius$ for 48 h under the
oxygen pressure of 115 atm, 175 atm and 280 atm, respectively. The
X-ray powder diffraction (XRD) was recorded at room temperature
using X Pert PRO X-Ray diffractometer  (Philips) with CuK$\alpha$
radiation ($\lambda$ = 1.5418 ${\rm \AA}$). The resistivity
measurements were performed using the standard ac four-probe
method. The magnetic field was supplied by a superconducting
magnet system (Oxford Instruments). Magnetization measurement was
carried out with a superconducting quantum interference device
(SQUID) magnetometer (MPMS-7XL, Quantum Design). We also
determined the oxygen content of the samples using
K$_2$Cr$_2$O$_7$ titration method. An appropriate amount of sample
(about 30 mg) is dissolved in the mixture of vitriol and phosphate
acid, then the high valent Co ions are deoxidized to divalent ones
with Fe$^{2+}$ ions, and finally the excess Fe$^{2+}$ ions are
titrated with K$_2$Cr$_2$O$_7$ solution. The oxygen content for
the as-fabricated samples was determined as 2.930$\pm$0.005. The
oxygen content increases after annealing under the oxygen
pressure, and changes to 2.959$\pm$0.005 at 115 atm,
2.978$\pm$0.003 at 175 atm and 2.990$\pm$0.002 at 280 atm,
respectively.

\section{EXPERIMENTAL RESULTS}

\subsection{Structural charactization}
In Fig. 1, the unit cell volumes of the as-fabricated
Pr$_{0.5}$Ca$_{0.5}$Co$_{1-x}$Fe$_x$O$_{3-\delta}$ ($x$ = 0, 0.05,
0.10 and 0.15) samples determined by  the XRD patterns at room
temperature are plotted against $x$. In the determination of the
lattice parameters, the orthorhombic structural symmetry is assumed
and the volume is estimated for the cell described by
$\sqrt{2}a_{\rm p} \times 2a_{\rm p} \times \sqrt{2}a_{\rm p}$
(space group Pnma) with $a_{\rm p}$ being the lattice constant of
the cubic perovskite cell. Such a crystallographic structure has
been used in Pr$_{1-x}$Ca$_{x}$CoO$_3$ system previously by other
anthors.\cite{Tsubouchi1,Tsubouchi2,Fujita} The volume increases
with $x$, because ionic radius of Fe$^{3+}$ is slightly larger than
that of Co$^{3+}$ and Co$^{4+}$.\cite{Shannon,Shannon1} The unit
cell volumes of the iron-free sample is also shown in Fig. 1 with
varying the annealing oxygen pressure.  The volume decreases with
increasing oxygen pressure (oxygen content), similar to the results
observed in La$_{0.5}$Sr$_{0.5}$CoO$_{3-\delta}$.\cite{Haggerty}

\subsection{The magnetic and transport properties for the
as-fabricated samples}

Figure 2 shows the temperature dependence of the zero field cooled
(ZFC) and field cooled (FC) molar magnetization $M$ at $H$ = 0.1 T
for the as-fabricated
Pr$_{0.5}$Ca$_{0.5}$Co$_{1-x}$Fe$_x$O$_{3-\delta}$ ($x$ =0, 0.05,
0.10 and 0.15) samples. It clearly shows that all samples undergo a
ferromagnetic transition at the $T_{\rm c}$ (taken at the inflection
point of the FC curve) which is enhanced from about 77 K to 86 K
with increasing $x$ from 0 to 0.15.The iron-doped samples exhibit
almost the same FC $M$ below 50 K, their molar magnetization is much
larger that that of the Fe-free sample. As shown in Fig.2, the
magnetic moment of the Fe-free sample at 4 K, is 2470 emu/mol
($\approx$ 0.44$\mu_{\rm B}$/Co), while it increases to 2950 emu/mol
($\approx$ 0.53$\mu_{\rm B}$/Co) by the Fe doping. From the $M-H$
loops collected at 4 K for these four samples (shown Fig. 3),
obvious spontaneous magnetization ($M_{\rm s}$) and clear hysteresis
are observed, consistent with the ferromagnetic transition. A
pronounced characteristics in Fig.3 is that the $M_{\rm s}$
increases with the iron doping level. In contrast to the sudden
enhancement of spontaneous magnetization with increasing $x$ from 0
to 0.05, a slight increase in the $M_{\rm s}$ is observed with
further increasing $x$,  in accord with the results observed in
Fig.2. Apparently, the Fe doping enhances the ferromagnetism.
Another marked feature can be found in Fig. 3 is that the three
iron-doped samples have the same coercive force of 2535 Oe while the
iron-free sample shows a much larger coercive force of 3475 Oe.

Figure 4 presents the resistivity, $\rho$, as a function of
temperature for the as-fabricated
Pr$_{0.5}$Ca$_{0.5}$Co$_{1-x}$Fe$_x$O$_{3-\delta}$ ($x$= 0, 0.05,
0.10 and 0.15) samples. The resistivity grows monotonically with
increasing the iron doping level. All samples shows insulating
behavior in the high temperatures. With decreasing temperature,
the resistivity for all samples exhibits the change of slope at a
temperature of $T_{\rm p}$, and a metal-insulator transition
occurs at $T_{\rm p}$ for the samples with $x$= 0 and 0.05. The
values of $T_{\rm p}$ are 74.3, 80.2, 82.0, and 85.7 for $x$ = 0,
0.0.5, 0.10 and 0.15, respectively. It is found that the $T_{\rm
p}$ is nearly the same as the $T_{\rm c}$. In addition, the
variation of $T_{\rm p}$ with Fe doping coincides with that of
$T_{\rm c}$. These suggests that the inflection of the resistivity
is induced by the reduction of the spin scattering due to the
ferromagnetic transition. In spite of the increase of $T_{\rm p}$
and $T_{\rm c}$, the conductivity decreases with increasing iron
doping level. An increase by only four times in $\rho$ was
observed at 4 K with increasing $x$ from 0 to 0.15, however, an
increase in $\rho$ by over seven orders at 77 K occurred with iron
doping from 0 to 0.15 in the manganites in
Nd$_{0.67}$Sr$_{0.33}$Mn$_{1-x}$Fe$_{x}$O$_3$.\cite{Chang} When a
magnetic field is applied, the resistivity is suppressed
dramatically at low temperatures as shown in Fig. 5. It is the
typical behavior in an itinerant ferromagnet and the resistivity
is reduced by magnetic field due to the suppression of
spin-scattering. The magnetoresistance
$\Delta\rho$/$\rho$=[$\rho(H)-\rho(0)$]/$\rho$ at $H$ = 6 T is
also shown in Fig. 5. Negative $\Delta\rho$/$\rho$ and a peak of
$\Delta\rho$/$\rho$ around $T_{\rm p}$ are observed for all the
samples. As $x$ increases, the magnitude of the peak is reduced
gradually. On the contrary, the magnitude of $\Delta\rho$/$\rho$
at lower temperatures is enhanced by increasing the iron content.
The decrease in magnitude of $\Delta\rho$/$\rho$ is induced from
the enhancement of ferromagnetism, because the reduction of
spin-scattering by the external magnetic field is smaller in a
stronger ferromagnet. Deduced from the increase of resistivity
with the Fe doping in the Fig. 4, it can be concluded that the Fe
doping in this cobalt oxide should induce disorder as well as the
enhancement of ferromagnetism, probably due to randomly
distribution of iron ions. This could be considered as the reason
of the increase in magnitude of $\Delta\rho$/$\rho$  at 4 K.

\subsection{The transport and magnetic properties for the
post-annealed samples}

Figure 6 displays the temperature dependence of the resistivity
for the Pr$_{0.5}$Ca$_{0.5}$Co$_{1-x}$Fe$_x$O$_{3-\delta}$ ($x$ =
0, 0.05, 0.10 and 0.15) samples after annealing at 600$\celsius$
under the oxygen pressure of 115 atm. It can be found that the
conductivity for the samples with $x$ =0.10 and 0.15 is enhanced ,
while the conductivity for the samples with $x$ = 0 and 0.05 is
reduced, relative to the as-fabricated samples. Especially, the
resistivity for the sample with $x$ = 0 exhibits a rapid increase
below about 75 K, an increase by almost 10 times in resistivity at
4 K relative to that at 75 K is observed. Similar behavior in the
iron-free sample have been observed in previous
reports\cite{Tsubouchi1,Tsubouchi2} and is attributed to the
occurrence of a transition from high-temperature IS state to
low-temperature LS state. However, the change in resistivity from
the transition point to 4 K ( about ten times ) is much less than
that observed in the reports (more than three orders in
magnitude).\cite{Tsubouchi1,Tsubouchi2} It is noted that such a
transition is induced after annealing under the high oxygen
pressure. Consequently, higher oxygen pressure annealing
procedures were carried out.

Figure 7 shows the resistivity as a function of temperature for
the Pr$_{0.5}$Ca$_{0.5}$Co$_{1-x}$Fe$_x$O$_{3-\delta}$ ($x$ = 0,
0.05, 0.10 and 0.15) samples after annealing at 600$\celsius$
under the oxygen pressure of 175 atm. Although the resistivity of
the sample with $x$ = 0.15 shows very slight change, the
resistivity for the samples with $x$ = 0.05 and 0.10 decreases
apparently relative to the samples annealing under the oxygen
pressure of 115 atm. Nevertheless, the temperature-dependent
behavior is almost unchanged. A strange thing can be found that
the temperature corresponding to the inflection of the resistivity
($T_{\rm p}$) remains unchanged with the annealing conditions.
This suggests that the ferromagnetic transition temperature for
the Fe-doped samples is independent with the variation of the
annealing conditions. Figure 7 indicates that the resistivity for
the iron-free sample increases by more than 30 times with
decreasing temperature from 70 K to 4 K, much larger than that in
the sample annealed under the oxygen pressure of 115 atm. This
confirms that annealing under the high oxygen pressure can enhance
the spin-state transition.

Although the spin-state transition has been manifested apparently
in the resistivity behavior in the iron-free sample after
annealing under the oxygen pressure of 175 atm, Figure 8 shows
that there is still no rapid drop in magnetization around the
spin-state transition temperature, which has been observed in
previous reports.\cite{Tsubouchi1,Tsubouchi2} Nevertheless, the
magnetization for the iron-free sample after annealing decreases
apparently compared to that shown in Fig. 2, indicating that the
magnetic moment per cobalt (0.24$\mu_{\rm B}$/Co at 4 K) decreases
dramatically. This is consistent with the occurrence of the
spin-state transition. Actually, the magnitude of $\rho$ above 75
K for the iron-free sample increases with increasing annealing
oxygen pressure (see Fig. 6 and Fig.7) could suggest the
depopulation of $e_{\rm g}$ electrons in the iron-free sample even
at room temperatures since the conduction is mainly dominated by
the transfer of $e_{\rm g}$ electrons. The iron-doped samples
still behave as a good ferromagnet with the same $T_{\rm c}$ as
that in Fig. 2. This is consistent with the results ($T_{\rm p}$)
indicated by the resistivity. However, one should note that the
magnetization at 4 K increases for the samples with $x$ = 0.10 and
0.15 after annealing compared to the as-fabricated samples, on the
contrary, the $M(T)$ for the sample with $x$ = 0.05 decreases.
This suggests that the degradation of the spin state in the sample
with $x$ = 0.05 is partly induced by annealing procedure under the
high oxygen pressure. Similar to the iron-free sample, the
resistivity of the sample with $x$ = 0.05 becomes larger than that
of the sample with $x$ = 0.10 after annealing under high oxygen
pressure (see Fig. 6, Fig. 7), and this is consistent with the
partly induced degradation of spin-state in the sample with $x$ =
0.05 by annealing procedure deduced from Fig. 8. The $M-H$ loops
collected at 4 K for these four samples are shown in Fig. 9. At
first glance, it is noted that the three iron doped samples
exhibits the same coercive force of 2410 Oe, similar to that
observed in the as-fabricated samples but slightly less. The
iron-free sample shows a smaller coercive force than those
observed in the iron-doped samples. The apparent spontaneous
magnetization and the clear hysteresis loop evidence the existence
of ferromagnetic order in all of these four samples. The
enhancement of $M_{\rm s}$ with increasing iron doping level is
observed, similar to that in Fig. 3. Much lower $M_{\rm s}$
(0.33$\mu_{\rm B}$) at 4 K is obtained in Fig. 9 for the iron-free
sample than that in Fig. 4, evidencing the degradation of
spin-state of cobalt ions and consistent with the results in Fig.
8.

In attempt to obtain the iron-free sample with an abrupt drop in
$M(T)$ at the spin-state transition temperature as observed in
literatures,\cite{Tsubouchi1,Tsubouchi2} further annealing
procedure under the oxygen pressure of 280 atm is performed.
Figure 10 shows the FC magnetization at $H$ = 0.1 T and the
resistivity at zero field as the function of temperature. Larger
enhancement in magnitude of resistivity from 75 K to 4 K is
obtained (compared to Fig. 6 and Fig. 7), at the same time, a cusp
can be observed in $M(T)$ curve at 66 K, corresponding to midpoint
of the sharp increase of resistivity as shown in Fig. 10.
Therefore, it can be concluded that annealing under the high
oxygen pressure promotes the spin-state transition in the
iron-free sample.

\section{DISCUSSIONS}

A most intriguing result in the present work is the enhancement of
the SST by the annealing procedure under the high oxygen pressure in
the iron-free sample. For the as-fabricated iron-free sample, there
is no obvious indication for a SST, while after annealing under the
oxygen pressure of 115 atm and 175 atm, the resistivity exhibits MIT
below 70 K, which is considered to be related to a SST, and at the
same time, the magnetization at low temperature is reduced
dramatically (see Fig. 11(a)). After annealing under the oxygen
pressure of 280 atm, a clear SST similar to that reported by other
authors\cite{Tsubouchi1,Tsubouchi2} is observed in the $M(T)$ curve.
We have demonstrated in Sec. ${\rm \@Roman{2}}$ that the oxygen
content is enhanced with increasing the oxygen pressure in the
annealing procedure. It is noted that the SST in
Pr$_{0.5}$Ca$_{0.5}$CoO$_{3-\delta}$ takes place in a system
including nearly 50 : 50 trivalent and tetravalent Co ions. At first
glance, one may argue that nearly 50 : 50 trivalent and tetravalent
Co ions be essential for the occurrence of the SST. However, Fujita
et al.\cite{Fujita,Fujita1} reported that the similar SST occurs in
(Pr$_{1-y}$R$^\prime_{y}$)$_{0.7}$Ca$_{0.3}$CoO$_3$ (R$^\prime$ =
Sm, Tb, Y). Therefore, the proportion of nearly 50 : 50 trivalent
and tetravalent Co ions is not necessary. As one knows, the spin
state of Co ions is mainly determined by the competition in
magnitude between $\Delta_{\rm CF}$ and $J_{\rm ex}$, which controls
the difference of the electronic energies, ${\Delta}E$, between the
spin states. Fig. 1 shows that the cell volume decreases with
increasing the oxygen content. The reduction of cell volume can be
induced by the decrease of the Co-O-Co angle or the shrink of the
Co-O band length. The former reduces the Co-Co transfer energy $t$
and the latter enhances the $\Delta_{\rm CF}$. Both effects favor to
stabilize a low-spin state. Therefore, the reduction of the cell
volume due to the increase of the oxygen content may be responsible
for the enhancement of SST by annealing procedure under the high
pressure oxygen.

It is apparent that the Fe doping in
Pr$_{0.5}$Ca$_{0.5}$CoO$_{3-\delta}$ suppresses the spin-state
transition. Even after annealing under the oxygen pressure of 175
atm, clear ferromagnetic transition is observed in the sample with
$x$ = 0.05, although the magnetization is reduced relative to that
in as-fabricated one (see Fig. 11 (b)). This suggests that the Fe
doping has the effect of destroying the mechanism for the
occurrence of SST. ${\Delta}E$ is dependent on the volume of the
CoO$_6$ octahedra and the Co-Co transfer energy $t$. The slight
expansion of the unit cell volume with the Fe doping is observed
(see Fig. 1). This could result from either the enlargement of the
CoO$_6$ octahedra or the decrease of the CoO$_6$ octahedra tilt,
i.e. the increase of the Co-O-Co angle. The former possibility
would reduce the $\Delta_{\rm CF}$ and the latter one would
enhance $t$. Both of these possible effects diminish the
${\Delta}E$ and suppress the tendency of spin-state transition.
Furthermore, Katsuki et al.\cite{Katsuki} presented that Fe-O
bonds are stronger than Co-O bonds. The strong covalency of Fe-O
bonds broadens the energy widths of the $e_{\rm g}$ bands through
competing with the Co-O bonds and consequently leads to a
narrowing of the energy gap $\Delta{E}$, which could enhance the
population of $e_{\rm g}$ electrons and stabilize an
intermediate-spin state.

The electrical transport is dramatically affected by the impact on
the spin-state transition from the annealing procedure under high
oxygen pressure and the Fe doping. In the as-fabricated samples,
the resistivity increases with increasing the Fe doping level in
spite of the enhancement of the ferromagnetism, which has been
attributed to the disorder induced by the Fe doping. While after
annealing under the high oxygen pressure of 115 atm, the evolution
of resistivity changes dramatically (see Fig. 6). Contrary to the
decrease of resistivity in the samples of $x$ = 0.10 and 0.15 due
to the increase of the carrier concentration (oxygen content), the
resistivity increases for the samples of $x$ = 0 and $x$ = 0.05
and becomes larger than that of the samples with higher Fe doping
level (at least at room temperature). As further annealing
procedure under 175 atm is performed, the magnitude of resistivity
of the sample with $x$ = 0.05 decreases and becomes to lie between
those of the samples with $x$ = 0.10 and 0.15, while that for the
iron-free sample keeps to increase still. As one knows, the
transfer of the $e_{\rm g}$ electrons determines the conduction in
the perovskite-type cobalt oxides. The enhancement of oxygen
content leads to the increase of carrier concentration, and this
causes the decrease of the resistivity with increasing annealing
pressure in the ferromagnetic samples with $x$ = 0.10 and 0.15.
While in the samples with $x$ = 0 and 0.05, besides the increase
of hole concentration, it has been pointed out that SST (partly
for $x$ = 0.05) is induced by the increase of oxygen content,
suggesting the depopulation of $e_{\rm g}$ electrons. This leads
to the increase of resistivity in these two samples. For the two
samples, SST is much stronger for the iron-free sample, so that
the resistivity keeps increasing with the increase of the
annealing oxygen pressure. While the SST in the sample with $x$ =
0.05 occurs only partly, so that the increase of carrier
concentration still can lower the resistivity, which makes the
resistivity of this sample firstly increases  after annealing
under 115 atm then decreases after further annealing under 175 atm
(see Fig. 6 and Fig. 7). Therefore, it can be turned out that the
corporate effects of the disorder from the Fe doping, the change
of carrier concentration, and spin-state transition lead to the
intriguing evolution of the resistivity with the annealing
procedures.

Another interesting result induced by the iron doping is the
enhancement of the ferromagnetism. As discussed above, an apparent
result of the Fe doping is the suppression of the SST. So a
possible candidate responsible for the enhancement of the
ferromagnetism may be the enhancement of spin state of Co ions,
which means the presence of more $e_{\rm g}$ electrons. However,
Fig. 11(b) indicates that although the magnetization for the
sample with $x$ = 0.05 decreases  with increasing the annealing
oxygen pressure, indicative of the existence of a partly induced
SST in this sample, the $T_{\rm c}$ remains unchanged before and
after annealing. For the samples with $x$ = 0.10 and 0.15 the
magnetization increases with increasing the annealing oxygen
pressure, but the $T_{\rm c}$ also remains unchanged before and
after annealing (See Fig. 2 and Fig. 8). Therefore, it suggests
that the rise of spin state due to the Fe doping could not be the
reason of the enhancement of the ferromagnetism. According to the
studies of M${\rm \ddot{o}}$ssbauer spectrum in the iron-doped
TbBaCo$_2$O$_{5.5}$ and
La$_{1-x}$Sr$_{x}$CoO$_{3}$,\cite{Kopcewicz,Czirak,Homanay} the
iron ions have the formation of Fe$^{3+}$ with high-spin $t_{\rm
2g}^3e_{\rm g}^2$ electronic configuration. If we assume the same
state of iron ions also exists in the present compositions, the
ferromagnetic superexchange between Fe$^{3+}$-O-Co$^{4+}$ in the
LS Co$^{4+}$ ions could be expected.\cite{Maignan} This would
enhance the ferromagnetism of system.

\section{CONCLUSION}
In summary, a detailed study of the magnetic and transport
properties in Pr$_{0.5}$Ca$_{0.5}$Co$_{1-x}$Fe$_x$O$_{3-\delta}$
(x = 0, 0.05, 0.10 and 0.15) polycrystalline bulks has been done.
Although no apparent structural change is associated with the iron
doping, ferromagnetism is systematically enhanced while
conductivity is suppressed for the as-fabricated samples.
Annealing under high pressure oxygen induces a spin state
transition in the iron-free sample, and the SST become more
obvious with increasing the oxygen pressure. Fe doping suppresses
the SST and enhances the ferromagnetic transition. The SST induced
by annealing under high oxygen pressure in the \emph{iron-free}
sample is attributed to the reduction of cell volume due to the
enhancement of the oxygen content. The enlargement of the cell
volume induced by the Fe doping and the stronger Fe-O bonds than
Co-O bonds are considered to lead to the suppression of SST in the
iron-doped samples. The existence of possible ferromagnetic
superexchange between Fe$^{3+}$-O-Co$^{4+}$ is taken into account
to understand the enhancement of ferromagnetism by the Fe doping.

\section{ACKNOWLEDGEMENT}
This work is supported by the grant from the Nature Science
Foundation of China and by the Ministry of Science
and Technology of China, and the Knowledge Innovation Project of Chinese Academy of Sciences.\\

\clearpage

\begin{figure}[htbp]
\centering\includegraphics[width=0.85\textwidth]{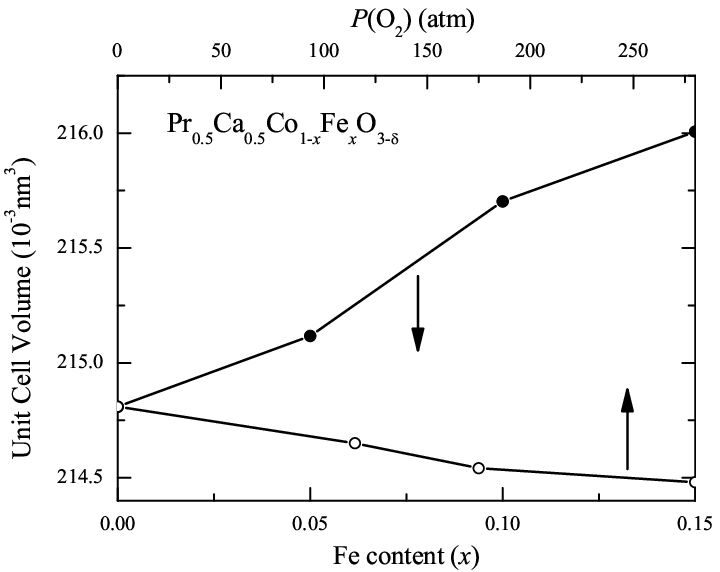}
\caption{The unit cell volume variation with $x$ for the
as-fabricated Pr$_{0.5}$Ca$_{0.5}$Co$_{1-x}$Fe$_{x}$O$_{3-\delta}$
polycrystals.}\label{Fig:Fig1}
\end{figure}

\begin{figure}[htbp]
\centering\includegraphics[width=0.85\textwidth]{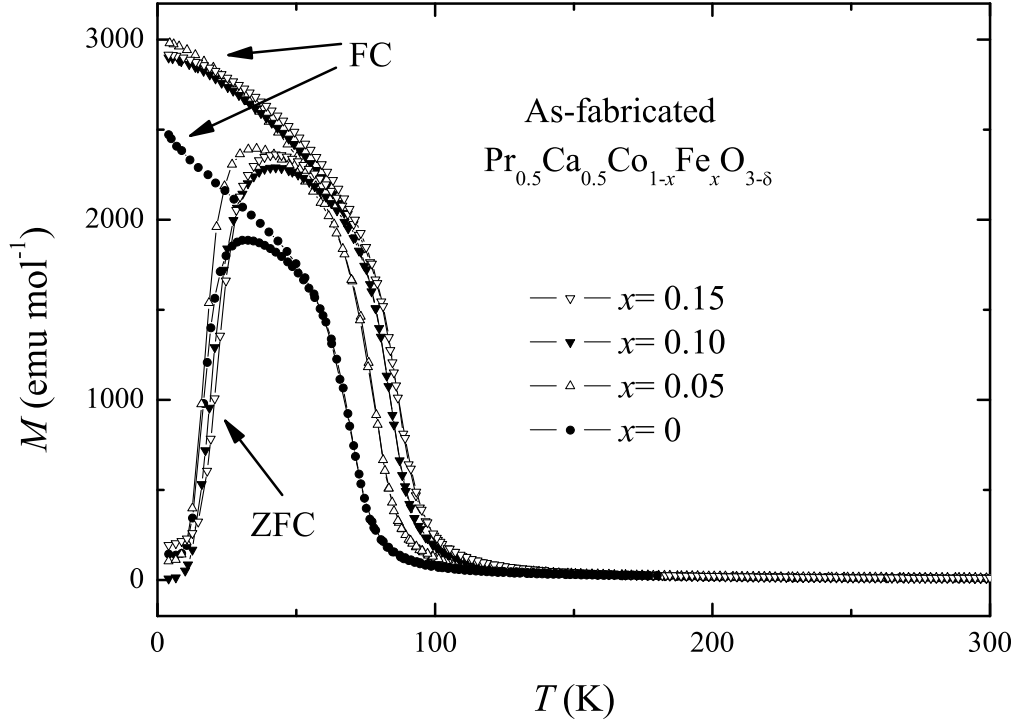}
\caption{The temperature dependence of the molar magnetization
recorded at $H$ = 0.1 T for the as-fabricated
Pr$_{0.5}$Ca$_{0.5}$Co$_{1-x}$Fe$_{x}$O$_{3-\delta}$
polycrystals.}\label{Fig:Fig2}
\end{figure}

\begin{figure}[htbp]
\centering\includegraphics[width=0.85\textwidth]{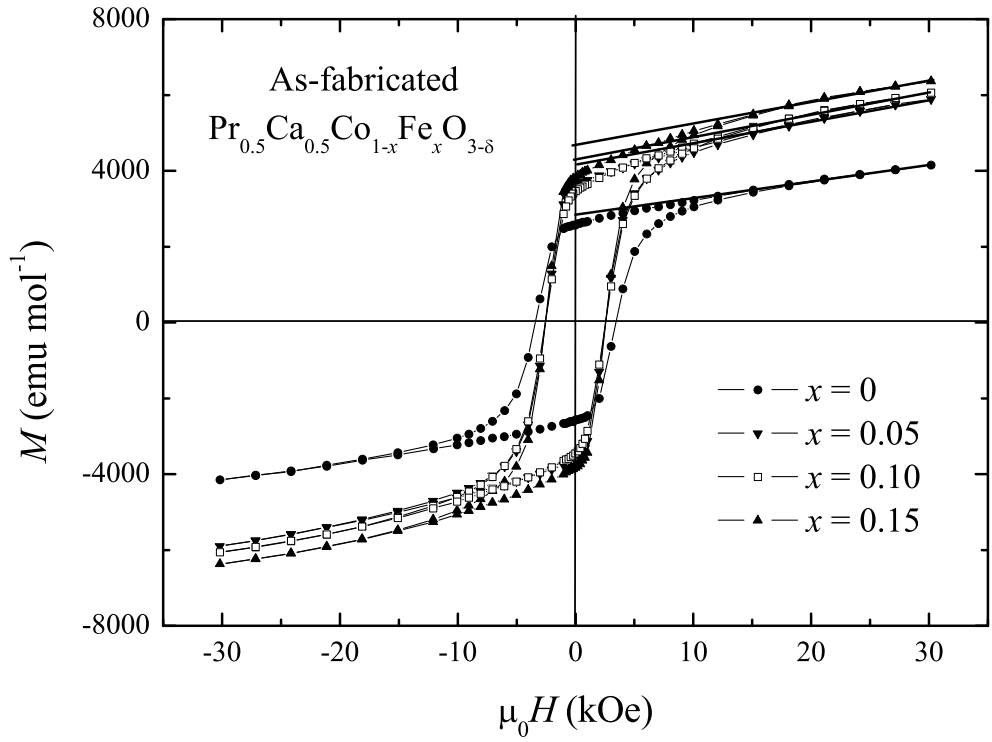}
\caption{The $M$-$H$ loop at 4 K for the as-fabricated
Pr$_{0.5}$Ca$_{0.5}$Co$_{1-x}$Fe$_{x}$O$_{3-\delta}$ polycrystals.
The solid lines extrapolate the $M(H)$ to $H$ = 0 to determined the
spontaneous magnetization.}\label{Fig:Fig3}
\end{figure}

\begin{figure}[htbp]
\centering
\includegraphics[width=0.85\textwidth]{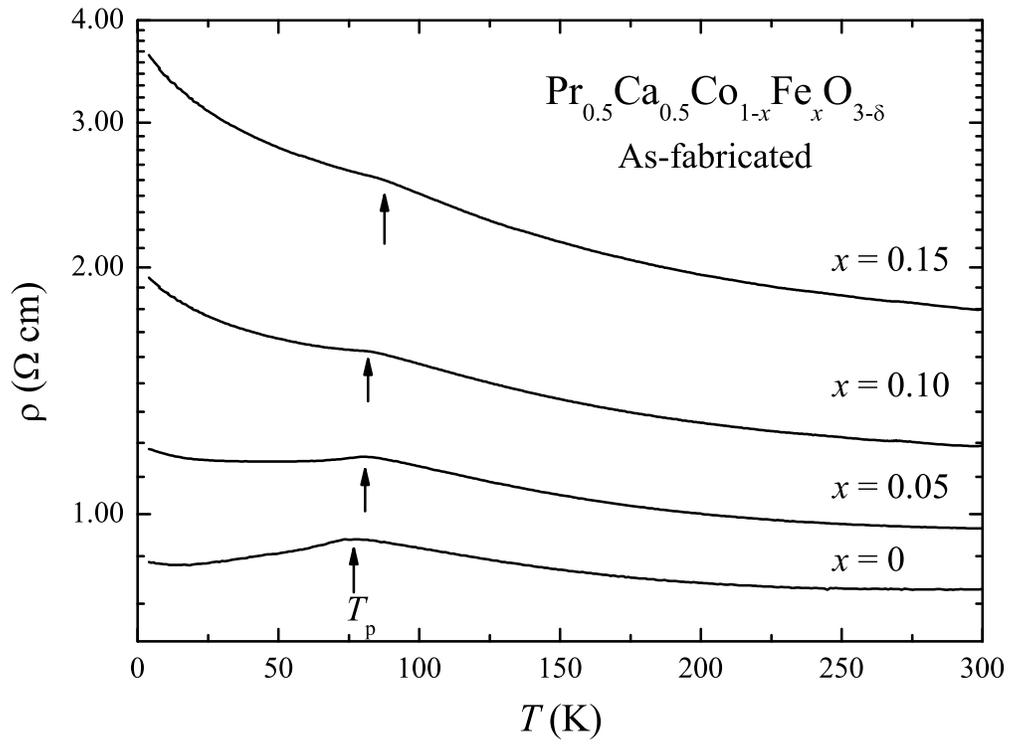}
\caption{The temperature dependence of the resistivity for the
as-fabricated Pr$_{0.5}$Ca$_{0.5}$Co$_{1-x}$Fe$_{x}$O$_{3-\delta}$
polycrystals.The dashed arrows point to the $T_{\rm
p}$.}\label{Fig:Fig4}
\end{figure}

\begin{figure}[htbp]
\centering
\includegraphics[width=0.85\textwidth]{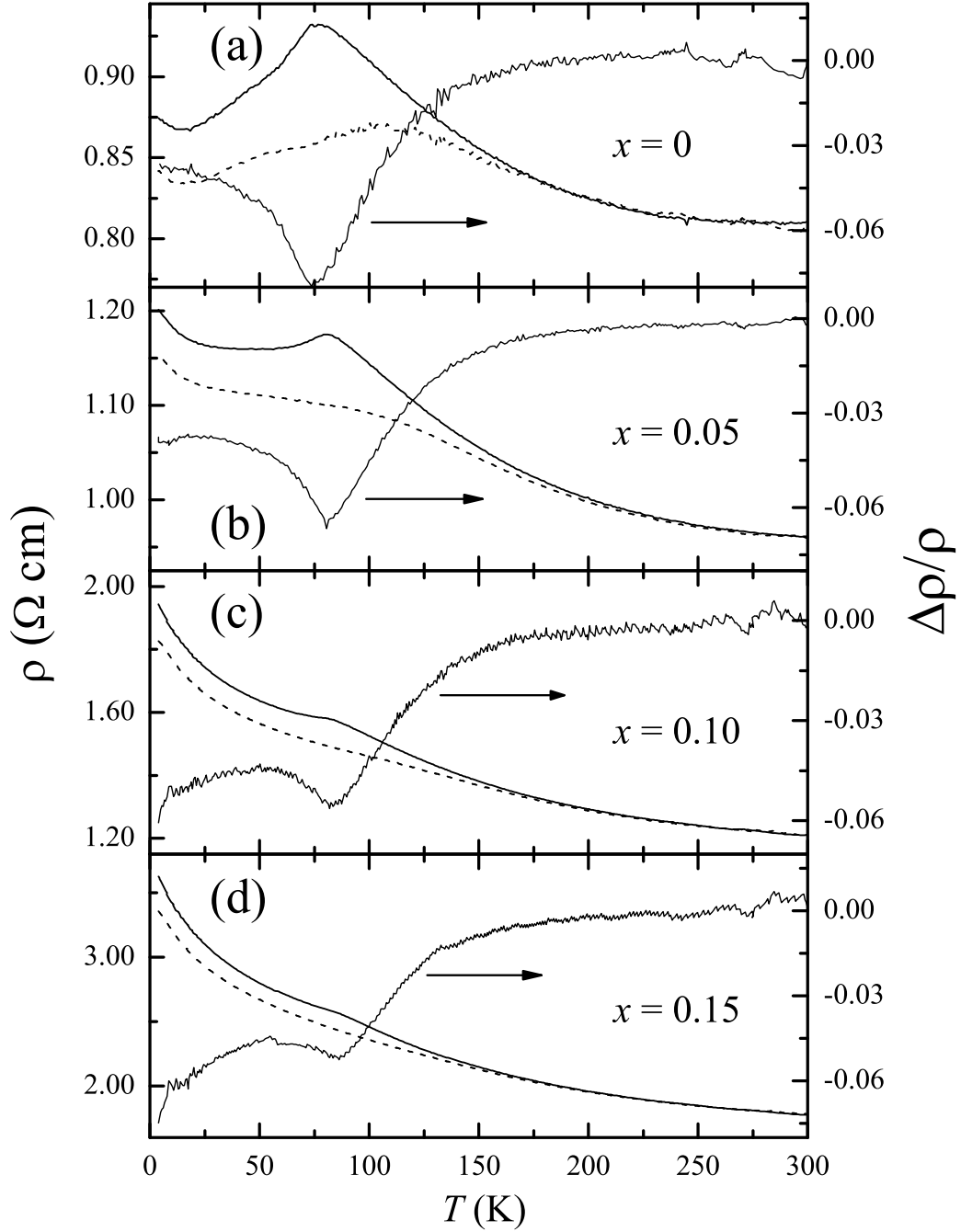}
\caption{Resistivity collected at $H$ = 0 (solid lines) and 6 T
(dash lines) and magnetoresistance ($\Delta\rho/\rho$) at $H$ = 6 T
as the function of the temperature for the as-fabricated
Pr$_{0.5}$Ca$_{0.5}$Co$_{1-x}$Fe$_{x}$O$_{3-\delta}$
polycrystals.}\label{Fig:Fig5}
\end{figure}

\begin{figure}[htbp]
\centering\includegraphics[width=0.85\textwidth]{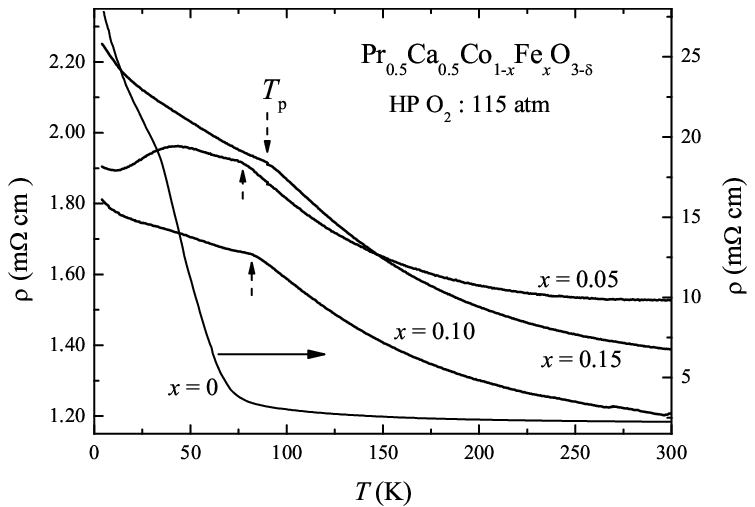}
\caption{The temperature dependence of the resistivity for the
Pr$_{0.5}$Ca$_{0.5}$Co$_{1-x}$Fe$_{x}$O$_{3-\delta}$ polycrystals
after annealing at 600$\celsius$ and under the oxygen pressure of
115 atm.The dashed arrows point to the $T_{\rm p}$.}\label{Fig:Fig6}
\end{figure}

\begin{figure}[htbp]
\centering\includegraphics[width=0.85\textwidth]{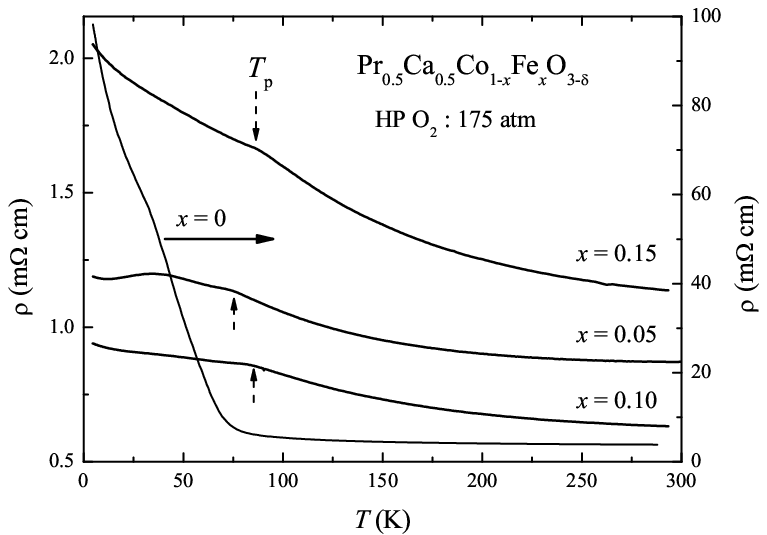}
\caption{The temperature dependence of the resistivity for the
Pr$_{0.5}$Ca$_{0.5}$Co$_{1-x}$Fe$_{x}$O$_{3-\delta}$ polycrystals
after annealing at 600$\celsius$ for 48 h under the oxygen pressure
of 175 atm. The dashed arrows point to the $T_{\rm
p}$.}\label{Fig:Fig7}
\end{figure}

\begin{figure}[htbp]
\centering\includegraphics[width=0.85\textwidth]{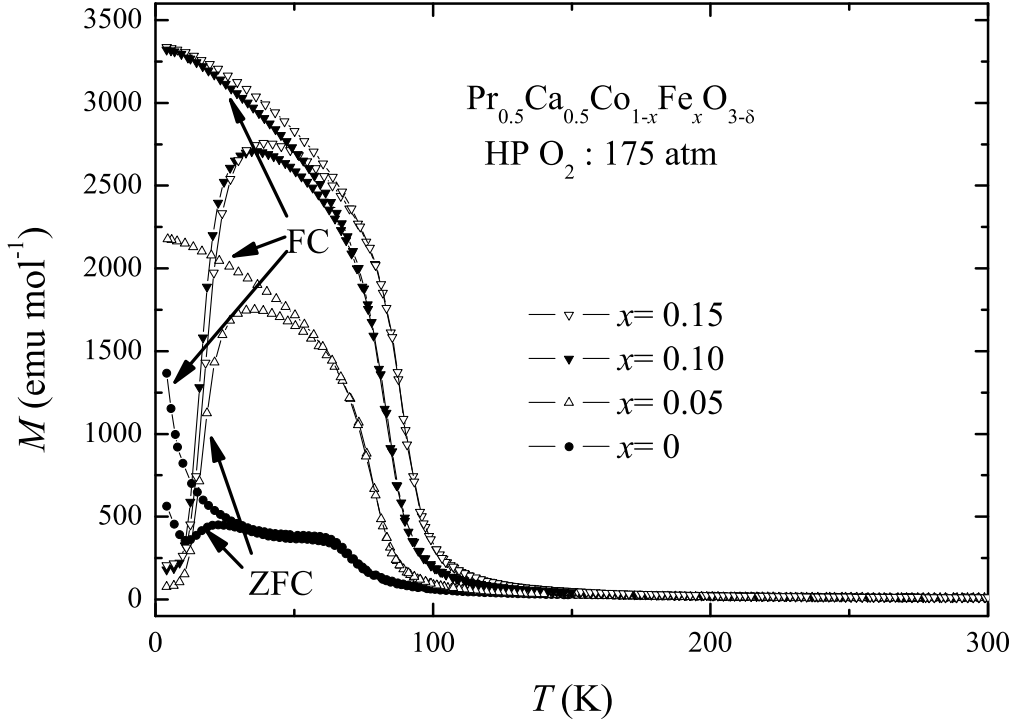}
\caption{The temperature dependence of the molar magnetization
recorded at $H$ = 0.1 T for the
Pr$_{0.5}$Ca$_{0.5}$Co$_{1-x}$Fe$_{x}$O$_{3-\delta}$ polycrystals
after annealing at 600$\celsius$ for 48 h under the oxygen pressure
of 175 atm.}\label{Fig:Fig8}
\end{figure}

\begin{figure}[htbp]
\centering\includegraphics[width=0.85\textwidth]{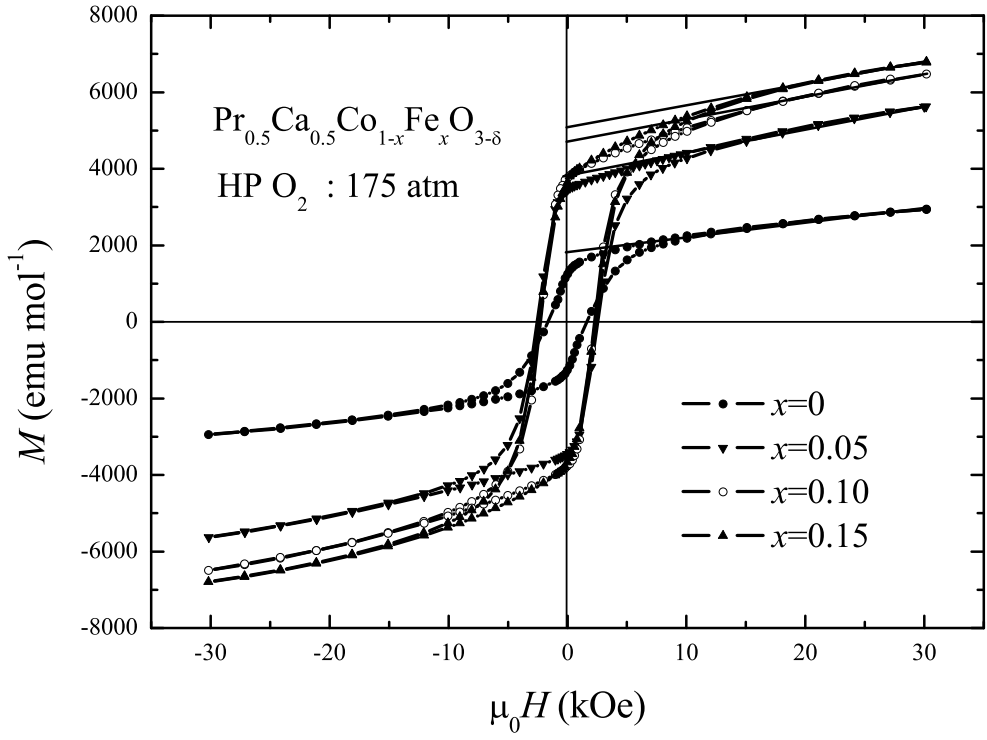}
\caption{$M-H$ loops at 4K for the
Pr$_{0.5}$Ca$_{0.5}$Co$_{1-x}$Fe$_{x}$O$_{3-\delta}$ polycrystals
after annealing at 600$\celsius$ for 48 h under the oxygen pressure
of 175 atm. The solid lines extrapolate the $M(H)$ to $H$ = 0 to
determined the spontaneous magnetization.}\label{Fig:Fig9}
\end{figure}

\begin{figure}[htbp]
\centering\includegraphics[width=0.85\textwidth]{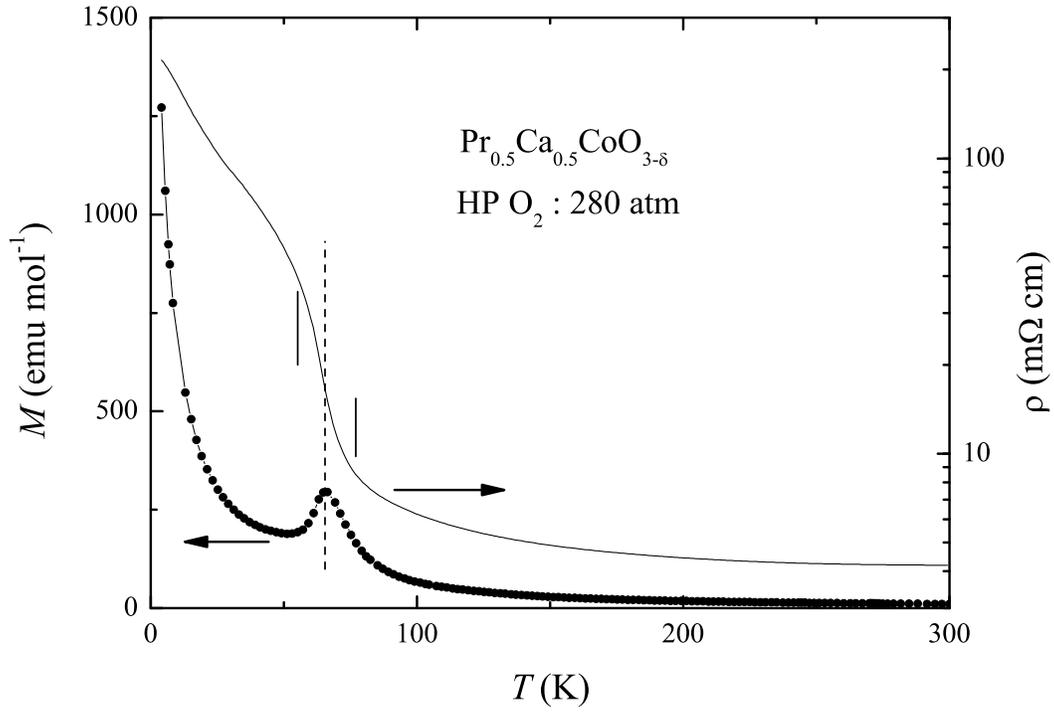}
\caption{The FC molar magnetization recorded at $H$ = 0.1 T and the
resistivity at zero field as the function temperature for the
Pr$_{0.5}$Ca$_{0.5}$CoO$_{3-\delta}$ polycrystals after annealing at
600$\celsius$ and under the oxygen pressure of 280
atm.}\label{Fig:Fig10}
\end{figure}

\begin{figure}[htbp]
\centering\includegraphics[width=0.85\textwidth]{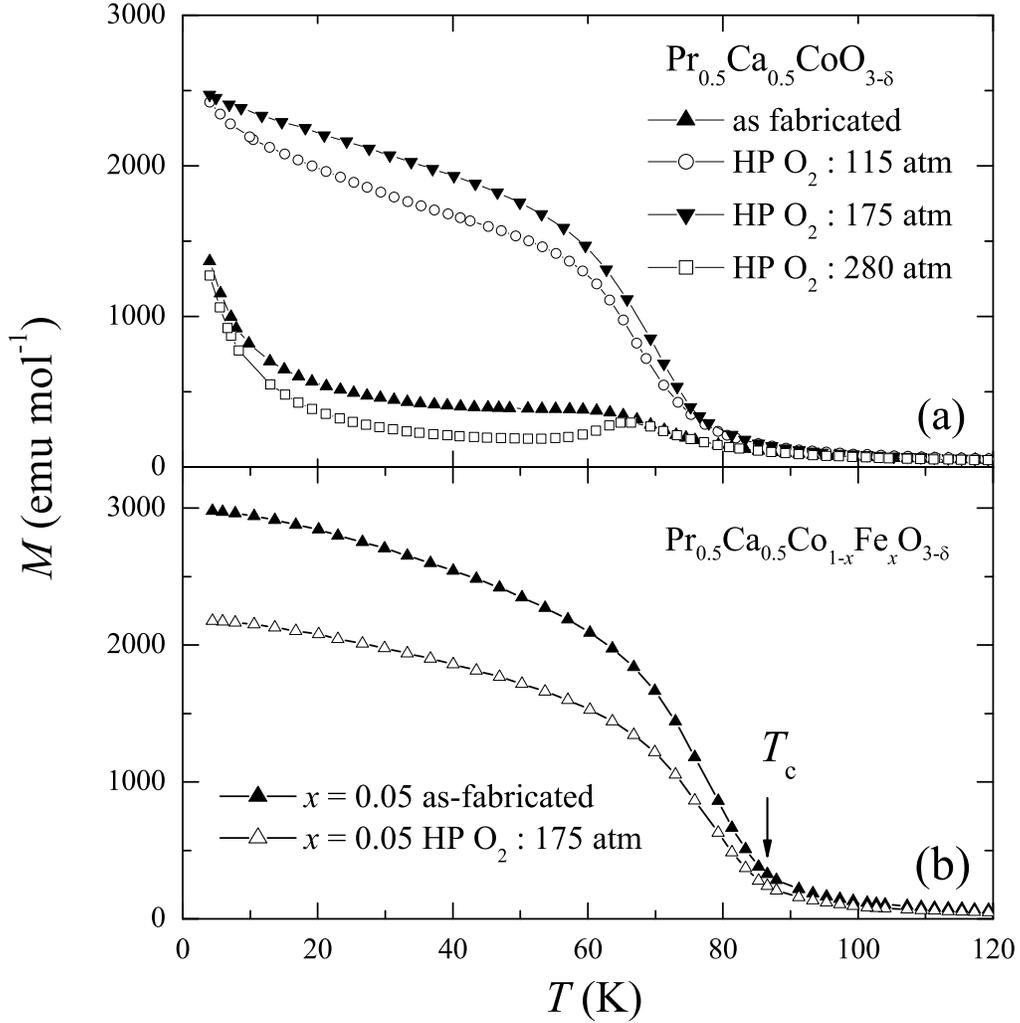}
\caption{(a)The FC molar magnetization at $H$ = 0.1 T as the
function temperature for the Pr$_{0.5}$Ca$_{0.5}$CoO$_{3-\delta}$
polycrystals obtained through different conditions.(b)The
temperature dependence of the FC molar magnetization at $H$ = 0.1 T
for the as-fabricated and the annealed
Pr$_{0.5}$Ca$_{0.5}$Co$_{0.95}$Fe$_{0.05}$O$_{3-\delta}$.}\label{Fig:Fig11}
\end{figure}

\end{document}